\documentclass[twocolumn,superscriptaddress,longbibliography,aps,pra,preprintnumbers]{revtex4-1}

\usepackage{graphicx}
\usepackage{bm}
\usepackage{color}
\usepackage{epstopdf}
\usepackage{amsmath}
\usepackage{amssymb}
\usepackage{epstopdf}

\usepackage[urlcolor=blue,colorlinks=true,citecolor=blue,
	linkcolor=blue,pdfstartview={FitH},bookmarks=false]{hyperref}

\usepackage{epsfig}
\usepackage[utf8]{inputenc}
\usepackage{dcolumn}% Align table columns on decimal point
\usepackage{bm}% bold math
\usepackage{hyperref}% add hypertext capabilities

\newcommand{\mb}[1]{ { \mbox{\boldmath{$#1$}}}  }

\begin{document}

\title{Quasiparticles of periodically driven quantum dot 
       coupled \\between superconducting and normal leads}

\author{Bartłomiej Baran}
 \email{bartlobaran@kft.umcs.lublin.pl}
\author{Tadeusz Domański}
 \email{doman@kft.umcs.lublin.pl}
\affiliation{Institute of Physics, M.\ Curie-Skłodowska University, 20-031 Lublin, Poland}

\date{\today}% It is always \today, today,

\begin{abstract}
We investigate subgap quasiparticles of a single level quantum dot coupled to 
the superconducting and normal leads, whose energy level is periodically driven 
by external potential. Using the Floquet formalism we determine the quasienergies 
and analyze redistribution of their spectral weights between individual harmonics 
upon varying the frequency and amplitude of the driving potential. We also
propose feasible spectroscopic methods for probing the in-gap quasiparticles 
observable in the differential conductance of the charge current averaged
over a period of oscillations.
\end{abstract}

\maketitle

\section{\label{sec:level1}Motivation}

Response of a quantum system on some abrupt quench \cite{quench} or periodically 
driven perturbations \cite{Polkovnikov-2011} can provide valuable insight into 
the dynamics of its quasiparticles and sometimes lead to emergence of novel phases 
without any analogy to equilibrium conditions \cite{floquet_matter}. Among the
prominent examples one can mention  such periodically driven phenomena, as: 
quantum time crystals \cite{Wilczek.2012}, topological insulators 
\cite{topological.insulators}, topological superconductors \cite{Klinovaja.2016}, 
zero and $\pi$ modes induced in the planar Josephson junctions \cite{Liu.2018} 
and many other. Such phenomena  affect the charge/spin 
transport through various heterostructures and might be promising 
for future applications. 

\begin{figure}
\includegraphics[height=45mm]{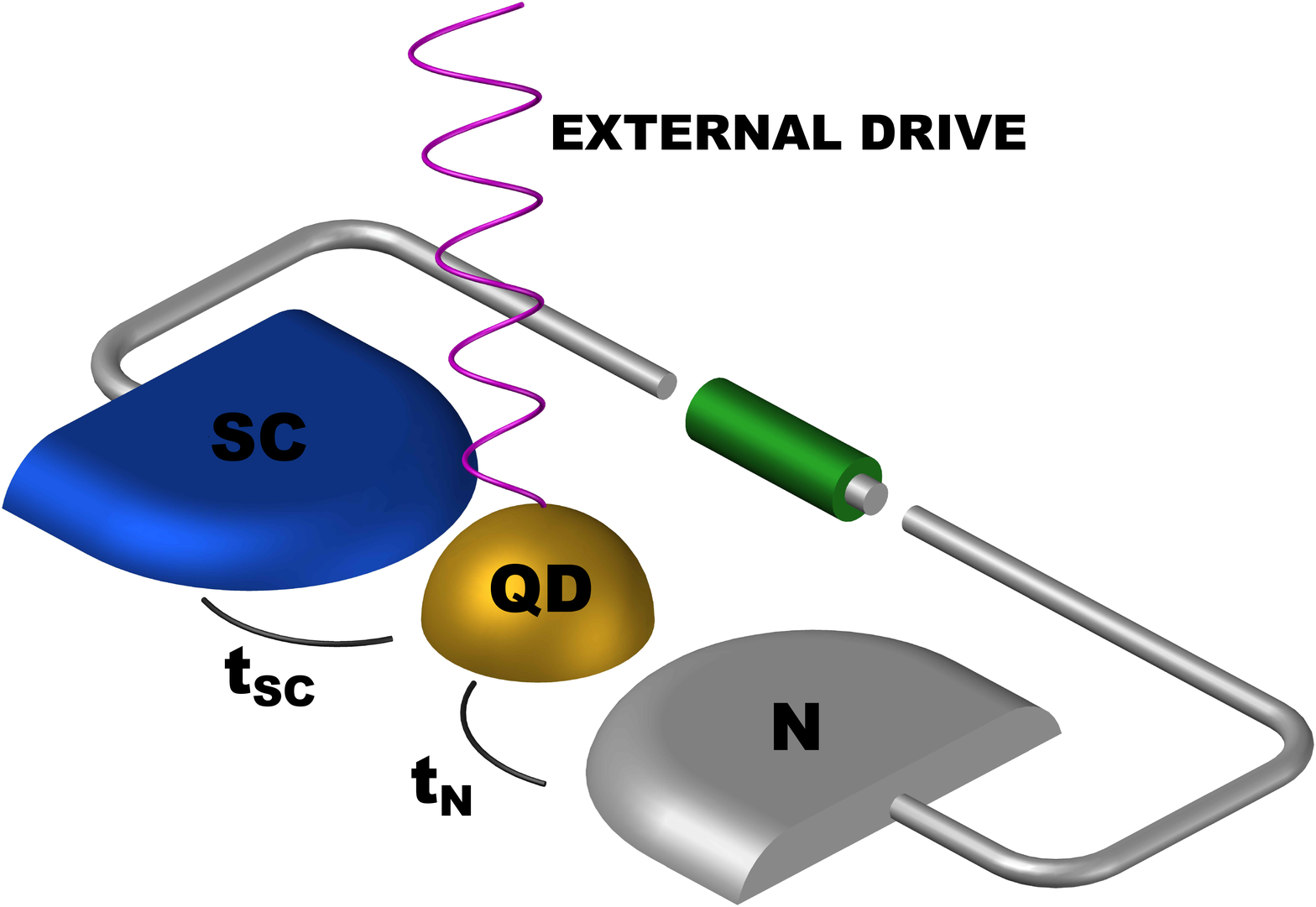}
\caption{Schematics of the externally driven quantum dot (QD) hybridized 
with superconducting (SC) and normal (N) electrodes by couplings $t_{SC}$ 
and $t_{N}$, respectively.}
\label{setup}
\end{figure}

In particular, very interesting effects arise at impurities embedded in 
superconducting reservoirs, where the bound (Andreev or Yu-Siba-Rusinov) 
states can appear in the subgap regime. Upon perturbing these impurities 
by some external periodic field they absorb or emit the field quanta, 
inducing the higher-order harmonic levels. Such features have been indeed 
reported experimentally \cite{PhysRevX.6.021014,PhysRevLett.115.216801}
but their detailed knowledge is far from clear. Since in-gap quasiparticles 
comprise the particle and hole ingredients, one may ask {\em whether the 
Andreev/Yu-Shiba-Rusinov states are going to split into a series of 
equidistant harmonics, or perhaps the normal harmonic quasienergies 
would undergo their internal splittings}. We investigate this issue 
here, considering the setup (Fig.~\ref{setup}) where the single level 
quantum dot is strongly coupled to the superconductor and weakly coupled 
to the normal lead. Energy level of this quantum dot can be periodically 
driven either by electromagnetic field or an alternating gate potential. 

Some aspects of the charge and heat transport through this setup has been 
recently discussed by L.~Arachea and R.~Rosa \cite{Arachea_Rosa.2018}, but 
specific nature of the quasienergies has not been addressed. 
Multiple in-gap features driven either by a.c.\ field \cite{Sun.1999} or 
monochromatic boson mode have been also discussed by several groups  
\cite{Timm.2012,Baranski_Domanski.2013,Bocian.2015,Cao.2017}. To our knowledge,
however, the frequency and the amplitude of external perturbations have not
been treated on equal footing. For this reason our purpose here is to 
study the subgap quasiparticles and their spectral weights, caused
by combined effect of the proximity-induced electron pairing and
external periodic perturbation.

The paper is organized as follows. We start by defining the microscopic 
model (Sec.\ \ref{sec:MODEL}) and next present methodological details
to treat the periodic driving (Sec.\ \ref{sec:METHOD}). Our main results 
are presented in Sec.\ \ref{sec:RESULTS}. Finally, in Sec.\ 
\ref{sec:conclusions}, we give a summary and brief outlook of open questions. 
Underlying ideas of the Floquet formalism are outlined in the Appendix.

%%%%%%%%%%%%%%%%%%%%%%%%%%%%%%%%%%%%%%%%%%%%%%%%%%%%%%%%%%%%%%%%%%%%%%%%%%%%%%%%%%%%%%%		%%%%%% MODEL
%%%%%%%%%%%%%%%%%%%%%%%%%%%%%%%%%%%%%%%%%%%%%%%%%%%%%%%%%%%%%%%%%%%%%%%%%%%%%%%%%%%%%%%	
\section{Microscopic model} 
{\label{sec:MODEL}

Setup comprising the quantum dot (QD) coupled to the normal (N) and 
superconducting (SC) reservoirs can be described by the Anderson 
impurity Hamiltonian
\begin{equation}
H(t) = H_{QD}(t)+H_{N}+H_{SC}+H_{TN}+H_{TS} .
\label{eq: Model Hamiltonian}
\end{equation}		
The time-dependence enters our setup through 
\begin{equation}
H_{QD}(t) = \sum_{\sigma} \varepsilon_d (t) d^{\dagger}_\sigma d_\sigma ,
\label{eq: Dot Hamiltonian}
\end{equation}		
where we assume periodic oscillations of the QD energy level $\varepsilon_d (t)=
\varepsilon_d + A \cos{(\omega t)}$. As usually, $d^{(\dagger)}_\sigma$ 
stands for the creation (annihilation) operator of the QD electrons with spin 
$\sigma =\{ \uparrow ,\downarrow \}$. Oscillations of the energy level 
$\varepsilon_d (t)$ are characterized by frequency $\omega$ and amplitude $A$. 
We assume, that they have no direct influence on electronic states of both 
external leads which are described by
\begin{eqnarray}
H_N &=& \sum_{k \sigma } \xi_{n k}  c^{\dagger}_{n k \sigma} c_{n k \sigma} ,
\label{eq: Normal lead Hamiltonian}
\\
H_{SC} &=& \sum_{k \sigma} \xi_{s k} c^{\dagger}_{s k \sigma} c_{s k \sigma} 
- \sum_{k} \left( \Delta c^{\dagger}_{s k \uparrow} c^{\dagger}_{s - k	\downarrow} 
+ \mbox{\rm h.c.}\right) .
\label{eq: Sc lead Hamiltonian}
\end{eqnarray}		
Here $c^{(\dagger)}_{\beta k \sigma}$ ($c^{(\dagger)}_{\beta k \sigma}$ ) are  
the creation (annihilation) operators of itinerant electrons with spin $\sigma$ 
and momentum $k$ in $\beta=N$  and $SC$ electrodes.  The energy gap of isotropic 
superconducting reservoir is denoted by $\Delta$. The energies $\xi_{\beta k}=\varepsilon_{\beta k} 
- \mu_{\beta}$ are measured with respect to the chemical potentials $\mu_{\beta}$, 
which can be detuned $\mu_{n}-\mu_{s}=eV$ by applying the bias $V$. 
The last terms of Hamiltonian (\ref{eq: Model Hamiltonian})
stands for hybridization of the QD with external leads 
\begin{equation}
H_{T\beta} = \sum_{k \sigma } \left( t_{\beta}  c^{\dagger}_{\beta k \sigma} 
d_{\sigma} + t^{*}_{\beta}  d^{\dagger}_{\sigma} c_{\beta k \sigma} \right) .
\label{hybridization}
\end{equation}
In what follows, we shall study the quasiparticle states appearing inside the
energy regime $|E | \leq \Delta $. For simplicity, we assume both hybridizations 
$t_{\beta}$ to be constant (momentum-independent).  

%%%%%%%%%%%%%%%%%%%%%%%%%%%%%%%%%%%%%%%%%%%%%%%%%%%%%%%%%%%%%%%%%%%%%%%%%%%%%%%%%%%%%%
\section{Methodology}	 
\label{sec:METHOD}
%%%%%%%%%%%%%%%%%%%%%%%%%%%%%%%%%%%%%%%%%%%%%%%%%%%%%%%%%%%%%%%%%%%%%%%%%%%%%%%%%%%%%%

Quantum systems described by the time-periodic Hamiltonians $H(t)=H(t+T)$,
where $T=2\pi/\omega$, can be treated within the Floquet formalism. Basic
ideas of this procedure are outlined in the Appendix. We extend this treatment 
onto the present setup, where the proximity induced on-dot pairing mixes 
the particle with hole degrees of freedom. We shall discuss below how to 
treat such effects in presence of the periodic driving.

The effective spectrum and transport properties of the N-QD-S setup can be obtained
using the Keldysh Green's function approach \cite{vanLeeuwen2006} combined 
with the Floquet technique \cite{PhysRevB.78.235124,PhysRevB.95.115303} to
account for the periodically oscillating QD level. 
Proximity effect induces pairing of the QD electrons, therefore
we introduce the matrix Green's functions in Nambu representation 
\begin{equation}
G^{\nu}_{d,d}(t,t')=  \left( \begin{tabular}{ c c }
$\langle \langle d_\uparrow (t) ; d^{\dagger}_\uparrow (t') 
\rangle  \rangle$ & $\langle \langle d_\uparrow (t) ; 
d_{\downarrow} (t')  \rangle\rangle $ \\
$\langle \langle d^{\dagger}_{\downarrow} (t) ; 
d^{\dagger}_\uparrow (t')   \rangle  \rangle$ & $\langle \langle
d^{\dagger}_{\downarrow} (t) ; d_{\downarrow} (t') \rangle \rangle$ 
\end{tabular} \right) ,
\label{eq: Keldysh Contour Nambu Green's Function}
\end{equation}	
where the upper index $\nu$ stands  either for the retarded ($\nu=r$), 
advanced ($\nu=a$) or Keldysh ($\nu=c$) functions. 
From the Heisenberg equation of motion one obtains 
\begin{equation}
G^{\nu}_{d,d}(t,t')=  g^{\nu}_{d,d}(t,t') 
+\int dt_{1}\sum_{k,\beta} g^{\nu}_{d,d}(t,t_{1}) t_{\beta}^{*} 
G^{\nu}_{\beta k, d}(t_{1},t') 
\label{eq:EOM}
\end{equation}	
where $g^{\nu}_{d,d}(t,t')$ is the (bare) Green's function of isolated 
QD, whereas $G^{\nu}_{\beta k, d \sigma}(t_{1},t')$ denotes the mixed 
function originating from hybridization of the QD  with itinerant 
electrons of external ($\beta=N$, $SC$) leads. Equation of 
motion for this mixed Green's function $G^{\nu}_{\beta k, d}(t_{1},t')$
yields the Dyson relation
\begin{eqnarray}
&&G^{\nu}_{d,d}(t,t') =  g^{\nu}_{d,d}(t,t') 
\label{eq:Dyson} \\ 
&+& \int dt_{1} \int dt_{2}
\sum_{\beta} g^{\nu}_{d,d}(t,t_{1}) {\mb \Sigma}^{\nu}_{\beta}(t_{1},t_{2})
G^{\nu}_{d, d}(t_{2},t') 
\nonumber
\end{eqnarray}	
with the selfenergy matrix
\begin{eqnarray}
{\mb \Sigma}^{\nu}_{\beta}(t_{1},t_{2}) = \sum_{k} \left| t_{\beta} \right|^{2} 
g^{\nu}_{\beta k, \beta k}(t_{1},t_{2}) . 
\label{selfenergies}
\end{eqnarray}	
The Green's functions and the selfenergies depend on two-time arguments
$t$ and $t'$, but such dependence can be substantially simplified owing 
to the discrete translational invariance $f(t,t')= f(t+nT,t'+mT)$  [where 
$n,m$ denote integer numbers] which holds in the steady limit that we 
are interested in.

Time periodicity can be conveniently treated, by transforming $t, t'$ to 
the relative $t-t'$ and average time $\left( t + t'\right)/2$ arguments  
and introducing the Wigner transformation \cite{PhysRevB.78.235124}. Here 
we follow slightly different convention \cite{Paaske_PhD}, introducing 
the transformation 
\begin{equation}
\begin{aligned}
&f_{nm}(\epsilon)=\int^{\infty}_{-\infty}dt'\frac{1}{T}\int^{T}_{0}dt 
e^{i(\epsilon+n\omega)t-i(\epsilon+m\omega)t'}f(t,t') 
\end{aligned}	
\label{eq: Fourier }
\end{equation}	
with the quasienergy $\epsilon$. Thereby we can recast time-convolutions 
appearing in (\ref{eq:EOM}) and in the Dyson equation (\ref{eq:Dyson}) 
by summations over the discrete harmonics $m,n$ and integral  over the first 
Floquet zone $\epsilon \in \left< -\omega/2 ; \omega/2 \right>$.

In the next step we diagonalize the bare Green's function $g^{-1}_{dd}(\varepsilon)$ 
with respect to its Floquet coordinates $n,m$ by the appropriate unitary matrix 
$\Lambda_{nl}(\varepsilon)=[\Lambda_{nl}(\varepsilon)]^{\dagger}$
\begin{equation}
\begin{aligned}
&\sum_{nm}\Lambda_{ln} (\epsilon) \left( g^{\nu}_{d,d}(\epsilon)\right)^{-1}_{nm}
\Lambda^{\dagger}_{ml} (\epsilon)= \left( Q^{\nu}_{d,d}(\epsilon)\right)^{-1}_{ll}.
\end{aligned}	
\label{eq:Diag1}
\end{equation}	
In this basis the retarded/advanced Green's function is simply expressed as
\begin{equation}
\begin{aligned}
\left( Q^{r,a}_{d,d}(\epsilon)\right)^{-1}_{ll}=\left( \epsilon +l\omega 
\pm i\eta^{+} \right) \mb{I} - \varepsilon^{0}_{d} \; \mb{\tau}_{z},
\end{aligned}
\label{Diag 2}	
\end{equation}
where $\mb{I}$ stands for identity matrix, $\mb{\tau}_{z}$ denotes $z$-component 
of the Pauli matrix, and $i\eta^{+}$ is an infinitesimal  positive imaginary value. 
We have chosen the time-dependent QD level $\varepsilon_{d}(t)$ of a cosine form, 
therefore the diagonalizing basis defined through (\ref{eq:Diag1}) is expressed 
by the Bessel functions of a first kind \cite{978-1-4020-6948-2}
\begin{eqnarray}
\Lambda_{nl}(\epsilon)&=&\frac{1}{T}\int_{0}^{T}dt e^{i(n-m)\epsilon t}
\; e^{-i\int_{0}^{t} dt' (\varepsilon_{d}(t')-\varepsilon_{d}(0))}
\nonumber \\
&=& J_{n-m}\left( \frac{A}{\omega} \right) .
\end{eqnarray}
Due to completeness of these Bessel functions, we can express
the bare Green's function in the following form 
\begin{equation}
\begin{aligned}
&\left( g^{r/a}_{d,d}(\epsilon) \right)_{nm} =\\
& \sum_{l} \left( \begin{tabular}{ c c }
$\frac{J_{n-l}(A/\omega)J_{m-l}(A/\omega)}{\epsilon \pm i\eta^{+}+l\omega-\varepsilon^{0}_{d}}$&$0$\\
$0$&$\frac{J_{n-l}(A/\omega)J_{m-l}(A/\omega)}{\epsilon \pm i\eta^{+}+l\omega+\varepsilon^{0}_{d}}$
\end{tabular} \right) .
\end{aligned}	
\label{eq: Diag 4}
\end{equation}
More detailed derivation of this transformation has been discussed 
in Refs \cite{PhysRevB.78.235124,Paaske_PhD}.
 
In the same way we express the selfenergies (\ref{selfenergies}) 
originating from hybridization of the QD with external leads
\begin{equation}
\left( \Sigma^{r/a}_{\beta}(\epsilon) \right)_{nm}=\sum_{k} |t_{\beta}|^{2} 
\left( g^{r/a}_{\beta k,\beta k}(\epsilon) \right)_{nm} .
\label{self energy}	
\end{equation}		
Since we are mainly interested in the subgap quasiparticles, we make use 
of the wide-band limit approximation \cite{PhysRevB.31.6207}, imposing 
the constant couplings $\Gamma_{\beta} \simeq 2\pi  |t_{\beta}|^{2} 
\rho(\mu_{\beta})$.  In the Floquet's space both the selfenergies become 
diagonal. The normal term is simply given as 
\begin{equation}
\begin{aligned}
&\left( \Sigma^{r/a}_{N}(\epsilon) \right)_{nm}=\mp  
 \left( \begin{tabular}{ c c }
$\frac{i\Gamma_{N}}{2}$ & $0$\\
$0$ & $\frac{i\Gamma_{N}}{2}$
\end{tabular} \right) \delta_{nm}
\end{aligned}
\label{SIGMA_N}	
\end{equation}		
whereas the superconducting contribution is non-diagonal in the Nambu 
representation \cite{Yamada.2011,Arachea_Rosa.2018} 
\begin{equation}
\left( \Sigma^{r/a}_{SC}(\epsilon) \right)_{nm} = -\frac{\alpha(\tilde{\varepsilon})
\; \Gamma_{SC}/2}{\sqrt{|(\tilde{\varepsilon}\pm i \eta^{+})^{2}-\Delta^{2}|}} \
 \left( \begin{tabular}{ c c }
$\tilde{\varepsilon}$ & $-\Delta$ \\
$-\Delta$ & $\tilde{\varepsilon}$
\end{tabular} \right)\delta_{nm},
\label{SIGMA_SC}	
\end{equation}	
where $\tilde{\epsilon}=\epsilon+ n\omega$ and $\alpha(\tilde{\epsilon})
=\Theta(\Delta-|\tilde{\epsilon}|)\pm isgn(\tilde{\epsilon})
\Theta(|\tilde{\epsilon}|-\Delta)$. The selfenergy (\ref{SIGMA_SC}) 
depends on the higher order harmonics $n\omega$ what has implications 
on the effective quasiparticle spectrum.

%%%%%%%%%%%%%%%%%%%%%%%%%%%%%%%%%%%%%%%%%%%%%%%%%%%%%%%%%%%%%%%%%%%%%%%%%%%%%%%%%%%%%%%	
\section{Effective spectrum}	 
\label{sec:RESULTS}
%%%%%%%%%%%%%%%%%%%%%%%%%%%%%%%%%%%%%%%%%%%%%%%%%%%%%%%%%%%%%%%%%%%%%%%%%%%%%%%%%%%%%%%	

In what follows we present some representative numerical results obtained 
for the periodically oscillating quantum dot, assuming $\varepsilon_{d}=0$,
$\Gamma_{N}=0.1\Gamma_{SC}$ and focusing on the zero temperature limit. 
Our main interest concerns the subgap quasiparticles and efficiency of 
the induced on-dot electron pairing. For this reason we start by discussing 
the superconducting atomic limit $\Delta \rightarrow \infty$ when the 
selfenergy (\ref{SIGMA_SC}) simplifies to its static value \cite{Rodero.2011}. 
Influence of the  energy gap $\Delta$ is discussed in Sec.\ \ref{finite_Delta}.

\subsection{In-gap quasiparticles} 
%\label{sec:Local density of states}
	
The effective QD spectrum driven by oscillations of the energy 
level $\varepsilon_{d}(t)$ can be characterized by the spectral 
function (diagonal in the Nambu space) defined as
\begin{equation}
\langle \rho_{d}(\epsilon) \rangle = \sum_{n} \left(  -\frac{1}{\pi} {\text{Im}}
\left[ G^{r}_{d,d}(\epsilon+i 0^{+}) \right]_{1,1} \right)_{nn} .
\label{DOS}	
\end{equation}		
Summation over the diagonal Floquet indices is here equivalent to 
averaging over the period $T$. For convenience we shall normalize this 
function (\ref{DOS}) multiplying it by $c=\frac{\pi}{2}\Gamma_{N}$. 
In the time-independent case ($A=0$ or $\omega=0$) this would imply, 
that $c \langle \rho_{d}(\epsilon) \rangle$ is equal to one 
for $\varepsilon$ coinciding with the subgap bound states. 

The normal QD (discussed in the Appendix) is characterized by 
a series of the harmonics $\varepsilon_{d}+n\omega$ (where $n$ 
stands for positive and negative integer numbers) whose spectral 
weights vary with the amplitude $A$. This structure changes qualitatively 
when the proximity induced on-dot pairing is taken into account. Fig.~\ref{fig:LDOS} 
shows the averaged spectral function \eqref{DOS} as a function of the quasienergy 
$\varepsilon$ and amplitude $A$ obtained for $\Gamma_{SC}/\omega=1$. We can 
notice, that  the normal quantum dot quasienergies $\varepsilon_{d}+n\omega$
split into the lower and upper branches.

\begin{figure}[ht]
\centering
\includegraphics[height=50mm]{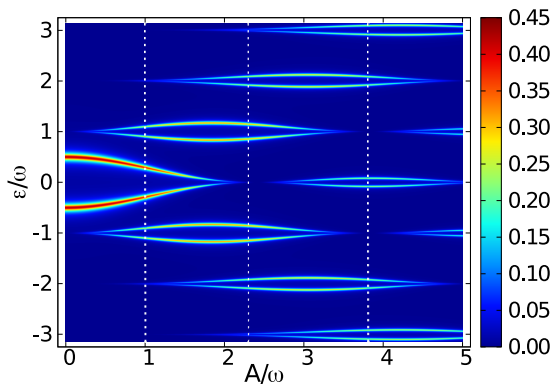}
\caption{The diagonal spectral function \eqref{DOS} of the quantum dot 
driven by periodic oscillations of its initial level $\varepsilon_{d}=0$, 
assuming $\Gamma_{SC}/\omega=1$ and  $\Gamma_{N}/\omega=0.1$.}
\label{fig:LDOS}
\end{figure}

Let us analyze this spectrum in more detail. For the stationary case 
the subgap spectrum consists of a pair of the Andreev bound states 
at $\pm \sqrt{\varepsilon_{d}^{2}+(\Gamma_{SC}/2)^{2}}$ \cite{Rodero.2011}. 
For our present configuration they acquire some finite
line-broadening (inverse life-time) originating from the coupling $\Gamma_{N}$ 
to a continuum of the normal lead electrons. Upon increasing the amplitude 
$\text{A}$ the quasiparticles branches (corresponding to $n=0$) 
gradually approach each other, and simultaneously the higher-order harmonics 
$|n|\geq 1$ are developed. Each of such higher-order quasiparticle branches 
does also reveal the splitting but its magnitude gets smaller and smaller 
with increasing $n$. The averaged spectrum (Fig.~\ref{fig:LDOS}) clearly 
displays, that such harmonics do not mix between themselves. They 
rather show up  {\em avoided crossing} behavior.  

Such variation of the quasiparticle energies with respect to $A$ is  
accompanied by considerable redistribution of their spectral weights. 
We observe that each of the harmonics gain and loose their weights 
upon varying the amplitude in roughly the same fashion as for the 
normal quantum dot (see Appendix). Fig.~\ref{fig:LDOSomega} illustrates 
the averaged spectral function versus the frequency $\omega$ of 
oscillations obtained for $A=2.2 \Gamma_{SC}$. Here we notice, that 
quasiparticle energies and ongoing transfer of their spectral weights 
between different harmonics  at larger frequencies produce the spectrum
comprising the higher order states near $\varepsilon_{d}+n\omega$ (like in the normal case) 
and one pair (of zero-th order) Andreev quasiparticles. 

\begin{figure}[h]
\centering
\includegraphics[height=50mm]{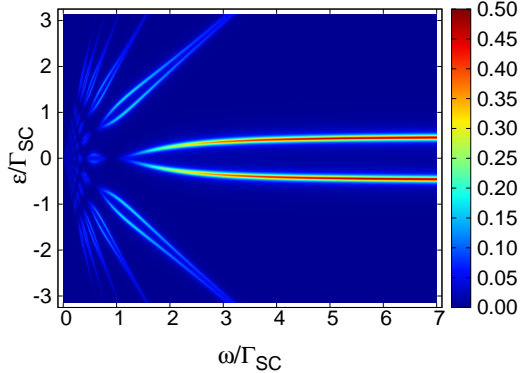}
\caption{Variation of the the averaged quasiparticle spectrum with respect 
to the frequency $\omega$ obtained for the constant  amplitude $A=2.2\Gamma_{SC}$, 
 assuming $\varepsilon_{d}{=}0$ and  $\Gamma_{N}{=}0.1\Gamma_{SC}$.}
\label{fig:LDOSomega}	
\end{figure}

\subsection{Induced on-dot pairing} 

To characterize the  induced on-dot pairing we introduce 
the off-diagonal (in Nambu space) spectral function 
\begin{equation}
\langle \rho_{off}(\epsilon) \rangle = \sum_{n} \left(  -\frac{1}{\pi} {\text{Im}}
\left[ G^{r}_{d,d}(\epsilon+i 0^{+}) \right]_{1,2} \right)_{nn} .
\label{offdiag_DOS}	
\end{equation}		
In Fig.~\ref{fig:offdiag} we show its variation with respect to the amplitude $A$. 
These quasiparticle branches are  reminiscent of the behavior shown 
in Fig.~\ref{fig:LDOS} for the diagonal spectral function. In the present case, 
however, the upper and lower branches in each harmonic have opposite signs.

\begin{figure}[ht]
\centering
\includegraphics[height=50mm]{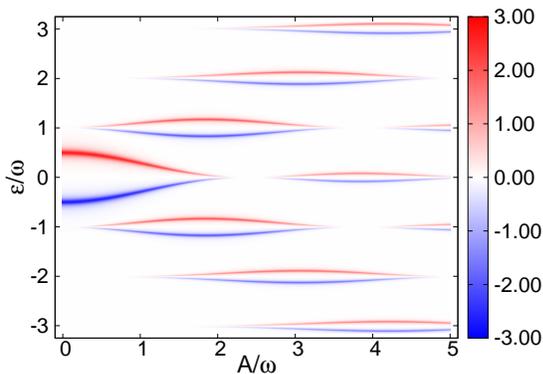}
\caption{The averaged off-diagonal spectral function (\ref{offdiag_DOS})
obtained for the same set of model parameters as in Fig.\ \ref{fig:LDOS}.}
\label{fig:offdiag}
\end{figure}

\begin{figure}[ht]
\centering
\includegraphics[height=50mm]{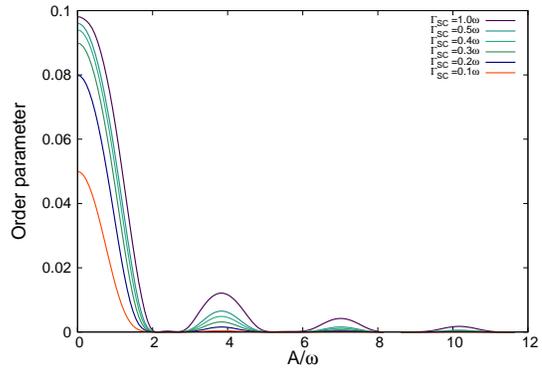}
\caption{Expectation value of the proximity induced on-dot pairing 
$\left< d_{\downarrow}d_{\uparrow}\right>$ versus the amplitude $A$  
obtained for $\varepsilon_{d}=0$, $\Gamma_{SC}/\omega=1$, $\Gamma_{N}/\omega=0.1$.}
\label{fig:order_parameter}
\end{figure}

We have also determined expectation value of the on-dot pairing potential 
$\left< d_{\downarrow}d_{\uparrow}\right>_{T}$ averaged over a period $T$. 
Its dependence on the amplitude $A$ is presented in Fig.~\ref{fig:order_parameter}.
This induced order parameter seems to be predominately sensitive to the amount
of spectral weight of the zero-th order harmonic states (it vanishes for 
such amplitude where the zero-level harmonic states loose their spectral 
weights). In the next section we shall check, whether the quasiparticle
spectrum and/or the induced on-dot pairing could be observable 
experimentally by the tunneling current measurements.

%%%%%%%%%%%%%%%%%%%%%%%%%%%%%%%%%%%%%%%%%%%%%%%%%%%%%%%%%%%%%%%%%%%%%%%%%%%%%%%%%%%%%%%
\subsection{Subgap charge current} 
{\label{sec:Charge current}
%%%%%%%%%%%%%%%%%%%%%%%%%%%%%%%%%%%%%%%%%%%%%%%%%%%%%%%%%%%%%%%%%%%%%%%%%%%%%%%%%%%%%%%

Spectrum of the QD spectrum can be probed experimentally only indirectly, through 
the transport properties. Let us briefly discuss  how to determine the time-dependent
charge current and its differential conductance. We focus on an adiabatic limit 
and use the Landauer's technique to describe the current induced in our setup 
by a small bias $V$, which detunes the chemical potentials $\mu_{N}=\mu_{SC}+eV$. 
To be specific, we assume the superconducting lead to be grounded $\mu_{SC}=0$. 

The  charge  current flowing from $\beta$-th electrode 
$I_{\beta}(t)=e\langle \dot{N}_{\beta}(t)\rangle$ can be expressed by  \cite{Sun.1999}
\begin{eqnarray}
I_{\beta}(t) &=& \frac{2e}{\hbar} \int dt_{1}  \text{Re} 
\left[ G^{r}_{d,d}(t,t_{1})\Sigma^{<}_{\beta}(t_{1},t) \right. \nonumber \\
&+& \left. G^{<}_{d,d}(t,t_{1})\Sigma^{a}_{\beta}(t_{1},t) \right]_{11-22} ,	
\label{time-dep-current}
\end{eqnarray}
where factor $2$ accounts for contributions from both spins whereas the diagonal 
elements \{11\} and \{22\} correspond to the particle and hole terms, respectively. 
In the Floquet's space  we can recast Eqn.\ \eqref{time-dep-current} 
to the form
\begin{eqnarray}
&&I_{\beta}(t)=\frac{2e}{\hbar}  \int_{-\omega /2}^{\omega /2} d\epsilon \sum_{n,m,p} 
\text{Re} \left\{ e^{-i(n-p)\omega t} \left[ \left( G^{r}_{d,d}(\epsilon)
\right)_{nm} \right. \right. \nonumber \\ 
&&\left. \left. \times \left( \Sigma^{<}_{\beta}(\epsilon)\right)_{mp} + \left( G^{<}_{d,d}(\epsilon) 
\right)_{nm} \left( \Sigma^{a}_{\beta}(\epsilon)\right)_{mp} \right]_{11-22}	\right\} .
\label{Time-MatrixCurrentFormula}
\end{eqnarray}

We have computed numerically the time-dependent current \eqref{Time-MatrixCurrentFormula} 
for several amplitudes $A$ marked by the dashed lines in Fig.\ \ref{fig:LDOS}. 
The current $I_N$ obtained for the bias voltage  $V=1\omega$ within 
a single period $T$ is displayed in Fig.~\ref{fig:time_dep_current}. For an 
opposite bias the symmetry relation $I_{N}(-V,t)=-I_{N}(V, t+\frac{T}{2})$
can be used. In general, we  hardly find any relevance of such time-dependent 
charge currents to effective quasiparticle spectrum of the driven quantum dot.

\begin{figure}[h]
\centering
\includegraphics[height=30mm]{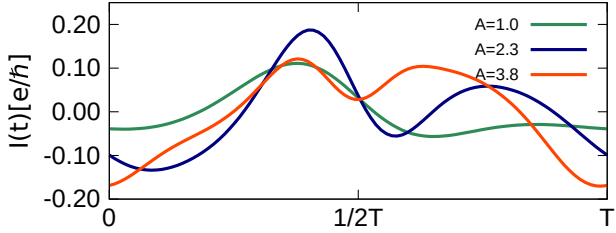}
\caption{Time-dependent current $I_N(t)$ obtained for $V=1\omega$, 
assuming $\varepsilon_{d}=0$ and the couplings $\Gamma_{SC}{=}1\omega$, $\Gamma_{N}{=}0.1\omega$.}
\label{fig:time_dep_current}
\end{figure}

In order to get
some correspondence with the effective QD spectrum let us 
analyze the transport properties averaged over the single period $T$. 
The averaged charge current can be obtained from (\ref{Time-MatrixCurrentFormula}) 
\begin{eqnarray}
\langle I_{\beta} \rangle &=&\frac{2e}{\hbar}  \int_{-\omega /2}^{\omega /2} d\epsilon \sum_{n,m} 
\text{Re} \left\{  \left[ \left( G^{r}_{d,d}(\epsilon)
\right)_{nm} \left( \Sigma^{<}_{\beta}(\epsilon)\right)_{mn} \right. \right. \nonumber \\ 
&+& \left. \left.   \left( G^{<}_{d,d}(\epsilon) 
\right)_{nm} \left( \Sigma^{a}_{\beta}(\epsilon)\right)_{mn} \right]_{11-22}	\right\} .
\label{averaged_current}
\end{eqnarray}
We  express the lesser Green's function $G^{<}_{d,d}(\epsilon)$ by a convolution 
of the retarded and advanced Green's function, using the selfenergy \cite{Sun.1999}
\begin{eqnarray}
\left( G^{<}_{\mu\nu}(\epsilon)\right)_{nm}
& = &\sum_{kl} \left[ \left( G^{r}_{\mu 1}(\epsilon) \right)_{nk} 
\left( \Sigma^{<}_{11}(\epsilon)\right)_{kl} 
\left( G^{a}_{1\nu}(\epsilon) \right)_{lm}
\right. \nonumber \\
&+& \left( G^{r}_{\mu 1}(\epsilon) \right)_{nk} 
\left( \Sigma^{<}_{12}(\epsilon)\right)_{kl} 
\left( G^{a}_{2\nu}(\epsilon) \right)_{lm} 
\nonumber \\
&+& \left( G^{r}_{\mu 2}(\epsilon) \right)_{nk} 
\left( \Sigma^{<}_{21}(\epsilon)\right)_{kl} 
\left( G^{a}_{1\nu}(\epsilon) \right)_{lm}
\nonumber \\
&+& \left. \left( G^{r}_{\mu 2}(\epsilon) \right)_{nk} 
\left( \Sigma^{<}_{22}(\epsilon)\right)_{kl} 
\left( G^{a}_{2\nu}(\epsilon) \right)_{lm} \right] ,
\label{G_lesser}	
\end{eqnarray}	
where $\mu,\nu {\in} \{1,2\}$. The lesser selfenergy matrix
\begin{equation}
\Sigma^{<}(\epsilon)= \Sigma^{<}_{N}(\epsilon)+\Sigma^{<}_{SC}(\epsilon)
\label{Sigma_lesser}	
\end{equation}	
can be given by
\begin{equation}
\left( \Sigma^{<}_{\beta}(\epsilon)\right)_{nm} = \left[ \left( \Sigma^{a}_{\beta}
(\epsilon)\right)_{nm} - \left( \Sigma^{r}_{\beta}(\epsilon)\right)_{nm} 
\right]f_{\beta}(\epsilon+n\omega),
\label{Sigma_lesser_Floquet}	
\end{equation}	
where $f_{\beta}(x)= 1/ \left[ e^{\left(x-\mu_{\beta}\right)/k_{B}T}+1 \right]$ 
is the Fermi-Dirac distribution function for electrons in $\beta$-th lead.

\begin{figure}[h]
\centering
\includegraphics[height=50mm]{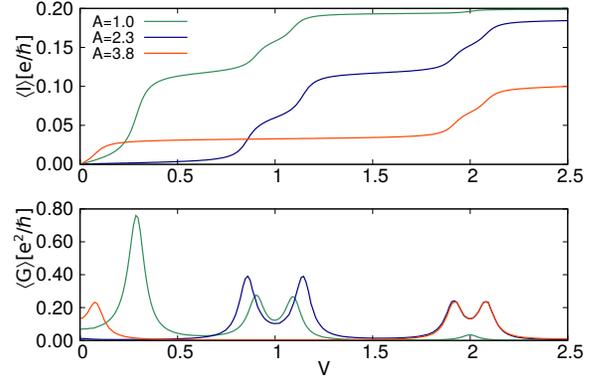}
\caption{The averaged current $\langle {I_N} \rangle$ and differential conductance  
$\langle {G_N} \rangle$ versu the applied bias voltage $V$ determined from
the Floquet's treatment for $\Gamma_{SC}{=}1\omega$, $\Gamma_{N}{=}0.1\omega$ 
and $\varepsilon_{d}{=}0$.}
\label{fig:average_current_plot}
\end{figure}

We have computed the averaged current given by Eqn.\ \eqref{averaged_current}
for the same set of parameters as discussed in Figs \ref{fig:LDOS} and \ref{fig:offdiag}.
Under equilibrium conductions the net current $\langle I_{\beta}\rangle$ vanishes, because 
incoming and outgoing charge transfers cancel each other. Fig.\ref{fig:average_current_plot} 
shows the averaged charge current (top panel) and its differential conductance 
(bottom panel) as functions of the applied voltage $V$ for three amplitudes
of the oscillations, as indicated. Enhancements of the differential 
conductance perfectly coincide with the energy dependent subgap quasiparticles 
(presented in Fig.\ \ref{fig:profile}) with the correspondence $\varepsilon \leftrightarrow eV$.

\begin{figure}[h]
\centering
\includegraphics[height=60mm]{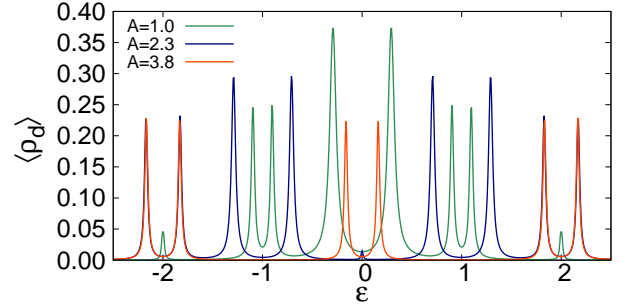}
\caption{Profiles of the diagonal spectral function for three amplitudes
of oscillations, as indicated. }
\label{fig:profile}
\end{figure}

Differential conductance of the charge current averaged over the period 
of oscillations would thus be able to experimentally probe the effective
quasiparticle spectrum, revealing the splittings of all harmonic levels.
	
\subsection{Finite \texorpdfstring{$\Delta$}{Lg} effects}
\label{finite_Delta}

In realistic situations the energy gap $\Delta$ is always finite, usually 
on the order of a few or fractions of meV. Let us inspect influence 
of such threshold on the effective quasiparticle spectrum. To be specific, we 
consider the  case  $\Delta = 0.5 \omega$ when the higher-order harmonics 
are pushed outside the superconducting energy gap window. 

\begin{figure}[h]
\centering
\includegraphics[height=50mm]{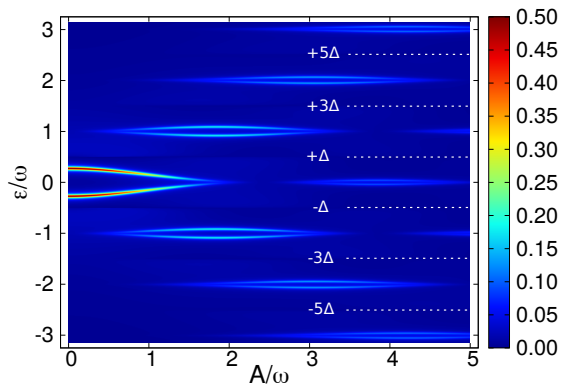}
\caption{Effective quasiparticle states obtained for $\Delta{=}0.5\omega$, 
assuming $\varepsilon_{d}=0$, $\Gamma_{SC}{=}1\omega$ and  $\Gamma_{N}{=}0.1\omega$.}
\label{fig:LDOS3}	
\end{figure}

Fig.~\ref{fig:LDOS3} presents the quasiparticle spectrum with respect to 
the varying amplitude $A$. In comparison to the limit $\Delta\rightarrow
\infty$, we notice that  outside the superconducting gap $\Delta$ the 
splitting of each harmonics substantially diminishes. This is rather well
expected behavior, but in addition we also observe further qualitative 
changes. When the amplitude $A$ exceeds the superconducting gap there 
occurs some partial leakage of the spectral weight towards the in-gap 
regime. It appears in a form of the continuous background,
corresponding to incoherent subgap states.

\begin{figure}[hb]
\centering
\includegraphics[height=50mm]{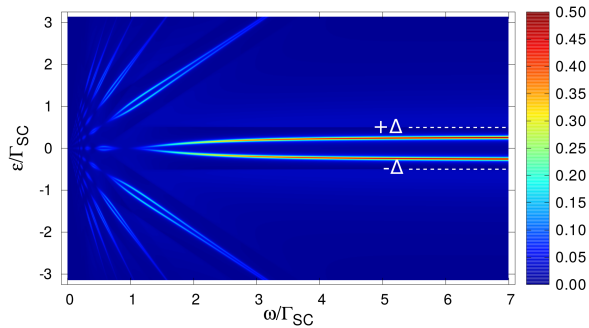}
\caption{Quasiparticle spectrum of the driven quantum dot obtained 
for the finite superconducting energy gap $\Delta=0.5\Gamma_{SC}$, 
assuming $\varepsilon_{d}{=}0$,  $\Gamma_{N}{=}0.1\Gamma_{SC}$ and
$\text{A}{=}2.2\Gamma_{SC}$.}
\label{fig:LDOSomega2}	
\end{figure}

Fig.\ \ref{fig:LDOSomega2} illustrates distribution of the spectral 
weight between the multiple harmonics, reveling their splittings and 
presence of the incoherent in-gap states. Let us notice, that for 
sufficiently fast oscillations we practically obtain the ordinary
(zero-level) Andreev quasiparticle states whereas all the rest 
of the spectrum is far outside the energy gap, arranged
into the higher order modes $\varepsilon_{d} \pm n \omega$. 
Close vicinity of the higher order harmonics is partly 
depleted from its continuous states -- this is exactly an opposite
tendency to the leakage of incoherent background displayed in
Fig.\ \ref{fig:LDOS3}. Finite value of the superconducting
energy gap is here manifested in quite new manner, without 
analogy to the stationary situations.

%%%%%%%%%%%%%%%%%%%%%%%%%%%%%%%%%%%%%%%%%%%%%%%%%%%%%%%%%%%%%%%%%%%%%%%%%%%%%%%%%%%%%%%		%%%%%% SUMMARY
%%%%%%%%%%%%%%%%%%%%%%%%%%%%%%%%%%%%%%%%%%%%%%%%%%%%%%%%%%%%%%%%%%%%%%%%%%%%%%%%%%%%%%%	

\section{Summary and outlook}	
\label{sec:conclusions}

We have studied effective spectrum of the single level quantum dot sandwiched 
between the superconducting and metallic electrodes and periodically driven by 
an external potential. We have analyzed variation of its quasienergies and spectral 
weights with respect to the frequency $\omega$ and the amplitude $A$ of oscillations. 
In stark contrast to the normal case (characterized by equidistant 
quasienergies $\varepsilon_{d}+l\omega$) we find, that the proximity induced 
electron pairing gives rise to the splitting of each harmonic level. Magnitude 
of such splitting is mostly pronounced in the zero-th harmonic state and 
gradually ceases for the higher harmonics. Distribution of the spectral 
weight between these split harmonic quasienergies is controlled by 
the amplitude to frequency ratio, roughly in the same fashion as for the 
normal case.

We have inspected the charge transport properties, establishing that
effective quasiparticle spectrum would be accessible via measurements of
the Andreev current averaged over a period of driven oscillations. 
Its differential conductance could verify, both the multi-harmonic 
quasiparticle energies, their internal splittings, and probe  
distribution of the spectral weights in each harmonic.  

We have also predicted  unusual (indirect) signatures of  
the superconducting energy gap $\Delta$ showing up in the quasiparticle 
spectrum. For sufficiently large amplitude of the oscillations (exceeding 
the energy gap $\Delta$ threshold) the subgap regime is poisoned by 
incoherent background states, corresponding to the short-time living 
quasiparticles. They emerge predominantly near such values of the amplitude 
to frequency ratio, where the spectral weight of the zero-th harmonic vanishes. 
This behavior goes hand in hand with suppression of the on-dot pairing, 
therefore it might be empirically detectable using 
the Josephson-type tunneling configurations.

We hope that verification of our predictions should be feasible 
with the presently available experimental techniques. Amongst 
important aspects unresolved in this paper let us point out the role 
of electron correlations. Interplay between the electron pairing and 
the local Coulomb repulsion might induce a changeover/transition 
of the ground state between the BCS-like singlet to the singly 
occupied doublet configuration. External driving potential might 
affect such phases in qualitatively different manner. This nontrivial 
issue, however, is beyond a scope of the present study and shall 
be addressed separately with use of appropriate many-body methods.

%%%%%%%%%%%%%%%%%%%%%%%%%%%%%%%%%%%%%%%%%%%%%%%%%%%%%%%%%%%%%%%%%%%%%%%%%%%%%%%%%%%%%%%

\section{ACKNOWLEDGMENTS}	 %\label{sec:ACKNOWLEDGMENTS}

We thank Jens Paaske for useful remarks. This work was supported by the National 
Science Centre (NCN, Poland) under grants UMO-2017/27/B/ST3/01911 (BB) and 
UMO-2018/29/B/ST3/00937 (TD).

%%%%%%%%%%%%%%%%%%%%%%%%%%%%%%%%%%%%%%%%%%%%%%%%%%%%%%%%%%%%%%%%%%%%%%%%%%%%%%%%%%%%%%%	

\appendix*
\section{Floquet formalism}
%\label{Floquet_technique}

Let us consider   time-dependent Hamiltonian $H(t)=H(t+T)$, where $T=2\pi / \omega$ is 
a period of external driving potential with the characteristic frequency  $\omega=2\pi/T$. 
Solution of the Schr\"{o}dinger equation can be formally represented by the Floquet's state 
$|\Psi_{\alpha} (t) \rangle = e^{-i\varepsilon_{\alpha} t}|\Phi_{\alpha} (t) \rangle$, 
where $|\Phi_{\alpha} (t) \rangle$ has the same periodicity $T$ as a perturbation. 
The wave-function  $|\Phi_{\alpha} (t) \rangle$ obeys the constraint $\left[ H(t)-
i\partial_{t}\right] |\Phi_{\alpha} (t) \rangle=\varepsilon_{\alpha}|\Phi_{\alpha}(t)\rangle$. 
with an eigenvalue $\varepsilon_{\alpha}$ \cite{PhysRevA.7.2203,PhysRev.138.B979}. 
In the specialistic literature $\left[ H(t)-i\partial_{t}\right]$ is dubbed quasioperator 
and $\varepsilon_{\alpha}$ quasienergy, respectively. 
Similarly to the Bloch treatment of translationally invariant spacial systems we can 
restrict to the interval $\varepsilon_{\alpha} \in \left[-\omega /2,\omega /2 \right)$, 
in analogy to the 1-st Brillouin zone. Performing the Fourier expansion 
of the eigen equation and we get 
\begin{equation}
\sum^{\infty}_{m=-\infty} (H_{nm}-n\omega\delta_{nm})|\Phi_{\alpha , m}\rangle 
=\varepsilon_{\alpha}|\Phi_{\alpha, n}\rangle,
\end{equation}	
where the Hamiltonian matrix elements are defined by $H_{nm}=\frac{1}{T}\int^{T}_{0}dt e^{i(n-m)\omega t}H(t)$ and 
the wave-function is $|\Phi_{\alpha,m}\rangle =\frac{1}{T}\int^{T}_{0}dt e^{in\omega t}|\Phi_{\alpha}(t)\rangle $. 
In the extended Hilbert space with time-independent Hamiltonian this can be written as 
$|\Psi_{\alpha} \rangle \rangle= \sum^{\infty}_{n=-\infty}|\Phi_{\alpha , n}\rangle \otimes |n \rangle$. 
Off-diagonal elements of the Hamiltonian matrix $H_{nm}$ correspond to transition amplitudes between 
the \textit{n}-th and \textit{m}-th Floquet's modes.

\begin{figure}[hb]
\centering
\includegraphics[height=50mm]{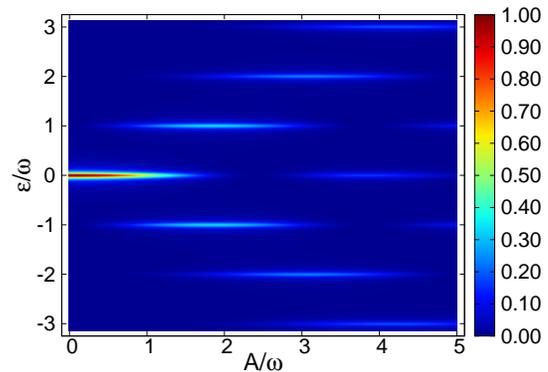}
\caption{Quasienergies of the normal quantum dot (coupled to 
metallic lead by $\Gamma_{N}/\omega=0.1$) appearing at $\varepsilon_{d}
\pm l \omega$ and variation of their spectral weights versus the
amplitude $A$  of oscillations.}
\label{fig:normalspectrum}	
\end{figure}

Fig.\ \ref{fig:normalspectrum} present the characteristic spectrum of a single level
quantum impurity driven by the periodic external potential of frequency $\omega$ and 
amplitude $A$. With increasing amplitude the initial level (here assumed to be
$\varepsilon_{d}=0$) is  replicated at higher harmonics $\varepsilon_{d}\pm l \omega$.
All these quasienergies are characterized by the spectral weights governed by 
the Bessel functions $J_{l}(A/\omega)$. They hence reveal, a kind of, oscillatory
variation with respect to $A$. Moreover, with an increasing amplitude the spectral
weight is shared between more and more harmonic states.

\bibliography{bibliography.bib}
 
\end{document}